\documentclass[12pt,preprint]{aastex}

\usepackage{epsfig}
\usepackage{graphicx}
\usepackage[dvips]{color}

\newcommand{\Fermic}{\textit{Fermi}}
\newcommand{\Fermi}{\Fermic\ }
\newcommand{\FermiLATc}{\Fermic\ LAT}
\newcommand{\FermiLAT}{\FermiLATc\ }

\shorttitle{Fermi Six}
\shortauthors{Furniss et al.}

\begin{document}

\title{VERITAS Observations\\of \\Six Bright, Hard-Spectrum \textit{Fermi}-LAT Blazars}

\author{
E.~Aliu\altaffilmark{1},
S.~Archambault\altaffilmark{2},
T.~Arlen\altaffilmark{3},
T.~Aune\altaffilmark{4},
M.~Beilicke\altaffilmark{5},
W.~Benbow\altaffilmark{6},
M.~B\"ottcher\altaffilmark{33},
A.~Bouvier\altaffilmark{4},
J.~H.~Buckley\altaffilmark{5},
V.~Bugaev\altaffilmark{5},
A.~Cesarini\altaffilmark{7},
L.~Ciupik\altaffilmark{8},
E.~Collins-Hughes\altaffilmark{9},
M.~P.~Connolly\altaffilmark{7},
W.~Cui\altaffilmark{10},
R.~Dickherber\altaffilmark{5},
C.~Duke\altaffilmark{11},
J.~Dumm\altaffilmark{12},
M.~Errando\altaffilmark{1},
A.~Falcone\altaffilmark{13},
S.~Federici\altaffilmark{14,15},
Q.~Feng\altaffilmark{10},
J.~P.~Finley\altaffilmark{10},
G.~Finnegan\altaffilmark{16},
L.~Fortson\altaffilmark{12},
A.~Furniss\altaffilmark{4,*},
N.~Galante\altaffilmark{6},
D.~Gall\altaffilmark{17},
S.~Godambe\altaffilmark{16},
S.~Griffin\altaffilmark{2},
J.~Grube\altaffilmark{8},
G.~Gyuk\altaffilmark{8},
D.~Hanna\altaffilmark{2},
J.~Holder\altaffilmark{18},
H.~Huan\altaffilmark{19},
P.~Kaaret\altaffilmark{17},
N.~Karlsson\altaffilmark{12},
Y.~Khassen\altaffilmark{9},
D.~Kieda\altaffilmark{16},
H.~Krawczynski\altaffilmark{5},
F.~Krennrich\altaffilmark{20},
K.~Lee\altaffilmark{5},
A.~S~Madhavan\altaffilmark{20},
G.~Maier\altaffilmark{14},
P.~Majumdar\altaffilmark{3},
S.~McArthur\altaffilmark{5},
A.~McCann\altaffilmark{21},
P.~Moriarty\altaffilmark{22},
R.~Mukherjee\altaffilmark{1},
T.~Nelson\altaffilmark{12},
A.~O'Faol\'{a}in de Bhr\'{o}ithe\altaffilmark{9},
R.~A.~Ong\altaffilmark{3},
M.~Orr\altaffilmark{20},
A.~N.~Otte\altaffilmark{23},
N.~Park\altaffilmark{19},
J.~S.~Perkins\altaffilmark{24,25},
A.~Pichel\altaffilmark{26},
M.~Pohl\altaffilmark{15,14},
H.~Prokoph\altaffilmark{14},
J.~Quinn\altaffilmark{9},
K.~Ragan\altaffilmark{2},
L.~C.~Reyes\altaffilmark{27},
P.~T.~Reynolds\altaffilmark{28},
E.~Roache\altaffilmark{6},
D.~B.~Saxon\altaffilmark{18},
G.~H.~Sembroski\altaffilmark{10},
D.~Staszak\altaffilmark{2},
I.~Telezhinsky\altaffilmark{15,14},
G.~Te\v{s}i\'{c}\altaffilmark{2},
M.~Theiling\altaffilmark{10},
S.~Thibadeau\altaffilmark{5},
K.~Tsurusaki\altaffilmark{17},
A.~Varlotta\altaffilmark{10},
V.~V.~Vassiliev\altaffilmark{3},
S.~Vincent\altaffilmark{14},
M.~Vivier\altaffilmark{18},
S.~P.~Wakely\altaffilmark{19},
T.~C.~Weekes\altaffilmark{6},
A.~Weinstein\altaffilmark{20},
R.~Welsing\altaffilmark{14},
D.~A.~Williams\altaffilmark{4},
B.~Zitzer\altaffilmark{29}
(The VERITAS Collaboration)\\
P.~Fortin\altaffilmark{6,*}, D.~Horan\altaffilmark{30,*}
and M.~Fumagalli\altaffilmark{31}, K.~Kaplan\altaffilmark{31} and J.~X.~Prochaska\altaffilmark{32}
}

\altaffiltext{*}{Corresponding authors: A.~Furniss: afurniss@ucsc.edu, P.~Fortin: pafortin@cfa.harvard.edu, D.~Horan: deirdre@llr.in2p3.fr}
\altaffiltext{1}{Department of Physics and Astronomy, Barnard College, Columbia University, NY 10027, USA}
\altaffiltext{2}{Physics Department, McGill University, Montreal, QC H3A 2T8, Canada}
\altaffiltext{3}{Department of Physics and Astronomy, University of California, Los Angeles, CA 90095, USA}
\altaffiltext{4}{Santa Cruz Institute for Particle Physics and Department of Physics, University of California, Santa Cruz, CA 95064, USA}
\altaffiltext{5}{Department of Physics, Washington University, St. Louis, MO 63130, USA}
\altaffiltext{6}{Fred Lawrence Whipple Observatory, Harvard-Smithsonian Center for Astrophysics, Amado, AZ 85645, USA}
\altaffiltext{7}{School of Physics, National University of Ireland Galway, University Road, Galway, Ireland}
\altaffiltext{8}{Astronomy Department, Adler Planetarium and Astronomy Museum, Chicago, IL 60605, USA}
\altaffiltext{9}{School of Physics, University College Dublin, Belfield, Dublin 4, Ireland}
\altaffiltext{10}{Department of Physics, Purdue University, West Lafayette, IN 47907, USA }
\altaffiltext{11}{Department of Physics, Grinnell College, Grinnell, IA 50112-1690, USA}
\altaffiltext{12}{School of Physics and Astronomy, University of Minnesota, Minneapolis, MN 55455, USA}
\altaffiltext{13}{Department of Astronomy and Astrophysics, 525 Davey Lab, Pennsylvania State University, University Park, PA 16802, USA}
\altaffiltext{14}{DESY, Platanenallee 6, 15738 Zeuthen, Germany}
\altaffiltext{15}{Institute of Physics and Astronomy, University of Potsdam, 14476 Potsdam-Golm, Germany}
\altaffiltext{16}{Department of Physics and Astronomy, University of Utah, Salt Lake City, UT 84112, USA}
\altaffiltext{17}{Department of Physics and Astronomy, University of Iowa, Van Allen Hall, Iowa City, IA 52242, USA}
\altaffiltext{18}{Department of Physics and Astronomy and the Bartol Research Institute, University of Delaware, Newark, DE 19716, USA}
\altaffiltext{19}{Enrico Fermi Institute, University of Chicago, Chicago, IL 60637, USA}
\altaffiltext{20}{Department of Physics and Astronomy, Iowa State University, Ames, IA 50011, USA}
\altaffiltext{21}{Kavli Institute for Cosmological Physics, University of Chicago, Chicago, IL 60637, USA}
\altaffiltext{22}{Department of Life and Physical Sciences, Galway-Mayo Institute of Technology, Dublin Road, Galway, Ireland}
\altaffiltext{23}{School of Physics and Center for Relativistic Astrophysics, Georgia Institute of Technology, 837 State Street NW, Atlanta, GA 30332-0430}
\altaffiltext{24}{CRESST and Astroparticle Physics Laboratory NASA/GSFC, Greenbelt, MD 20771, USA.}
\altaffiltext{25}{University of Maryland, Baltimore County, 1000 Hilltop Circle, Baltimore, MD 21250, USA.}
\altaffiltext{26}{Instituto de Astronomia y Fisica del Espacio, Casilla de Correo 67 - Sucursal 28, (C1428ZAA) Ciudad Aut—noma de Buenos Aires, Argentina}
\altaffiltext{27}{Physics Department, California Polytechnic State University, San Luis Obispo, CA 94307, USA}
\altaffiltext{28}{Department of Applied Physics and Instrumentation, Cork Institute of Technology, Bishopstown, Cork, Ireland}
\altaffiltext{29}{Argonne National Laboratory, 9700 S. Cass Avenue, Argonne, IL 60439, USA}
\altaffiltext{30}{Laboratoire Leprince-Ringuet, \'Ecole polytechnique, CNRS/IN2P3, Palaiseau, France}
\altaffiltext{31}{Department of Astronomy and Astrophysics, University of California, 1156 High Street, Santa Cruz, CA 95064}
\altaffiltext{32}{Department of Astronomy and Astrophysics, UCO/Lick Observatory, University of California, 1156 High Street, Santa Cruz, CA 95064}
\altaffiltext{33}{Astrophysical Institute, Department of Physics and Astronomy, Ohio University, Athens, OH 45701, USA}

\begin{abstract}
We report on VERITAS very-high-energy (VHE; E$\ge$100 GeV) observations of six blazars selected from the \textit{Fermi} Large Area Telescope First Source Catalog (1FGL). 
The gamma-ray emission from 1FGL sources was extrapolated up to the VHE band, taking gamma-ray absorption by the extragalactic background light into account.  This allowed the selection of six bright, hard-spectrum blazars that were good candidate TeV emitters.  Spectroscopic redshift measurements were attempted with the Keck Telescope for the targets without Sloan Digital Sky Survey (SDSS) spectroscopic data.  No VHE emission is detected during the observations of the six sources described here. Corresponding TeV upper limits are presented, along with contemporaneous \textit{Fermi} observations and non-concurrent \textit{Swift} UVOT and XRT data. The blazar broadband spectral energy distributions (SEDs) are assembled and modeled with a single-zone synchrotron self-Compton model.  The SED built for each of the six blazars show a synchrotron peak bordering between the intermediate- and high-spectrum-peak classifications, with four of the six resulting in particle-dominated emission regions.
\end{abstract}


\section{Introduction}

Blazars are active galactic nuclei (AGN) with a relativistic jet pointed close to the Earth line of sight.  AGN are thought to be powered by accretion onto a supermassive black hole ($\sim10^9$M$_{\sun}$) at the center of the host galaxy and are characterized by a double-peaked spectral energy distribution (SED) in the $\nu$F$_{\nu}$ representation.

The lower energy peak of the broadband SED is attributed to synchrotron emission of highly relativistic electrons and positrons in the presence of a tangled magnetic field.  In leptonic models, the higher energy peak is produced via the inverse-Compton (IC) up-scattering by the relativistic leptons of the synchrotron photons (synchrotron self-Compton, SSC) or a photon field external to the jet (external Compton, EC).  More details regarding leptonic non-thermal emission of blazars can be found in \cite{dermer, maraschi, sikora} and the references therein.  Alternative models associate the higher-energy peak to interaction of relativistic protons with an ambient photon field \citep{aharonian2002, bednarek, dar, mannheim, mucke, pohl}, or a hybrid population comprised of both leptons and hadrons \citep{markusHybrid}.

The blazar population is divided into two subclasses: flat spectrum radio quasars (FSRQs) and BL Lacs.  FSRQs are, to first order, more distant, more luminous, and have stronger emission lines than BL Lacs.  Historically, BL Lacs have been sub-classified based on their radio and X-ray flux ratios as low-, intermediate- or high-frequency-peaked BL Lac objects (LBL, IBL and HBLs, respectively) as described in \cite{padovani} and \cite{bondi}.  More recently, a classification based on the location of the synchrotron peak in frequency space ($\nu_{synch}$) has been proposed by \cite{abdoSED}, with low-spectrum-peaked (LSP) BL Lac objects having $\nu_{synch}$ below $10^{14}$\,Hz, intermediate-spectrum-peaked (ISP) BL Lac objects peaking between $10^{14}$ and $10^{15}$\,Hz and high-spectrum-peaked (HSP) BL Lac objects showing a peak above $10^{15}$ Hz.  HSP BL Lacs are the most common extragalactic object to be detected at very-high-energies (VHE; $E\ge100$ GeV), comprising 33 of 41 VHE BL Lac objects detected as of June 2012.  There have also been 4 ISPs \citep{acciari3C66A, acciariWComae, acciariPKS1424, ong1ES1440} and 4 LSPs \citep{mazin0716, mariotti, hoffman, albertBLLac} detected since the advent of VHE gamma-ray astronomy in the late 1960s.  In addition, three FSRQs, three radio galaxies, and two starburst galaxies complete the catalog of associated extragalactic objects \citep{tevcat}\footnote{\tt http://tevcat.uchicago.edu/}.

The potential scientific impact of increasing the catalog of VHE emitting extragalactic objects is substantial.  
A significant fraction of the power released from these objects is within the VHE band.  This makes the measurement of VHE blazar spectra an important component of the overall understanding of these objects.  With a better sample of well-measured VHE blazar spectra available for study, a population-based investigation of gamma-ray production in these objects through broadband SED modeling will be possible, providing means to answer the long-standing question of whether VHE gamma-ray emission results from leptonic or hadronic processes in AGN jets.  We can also apply the model-inferred properties of these new discoveries to understanding how the gamma-ray production may differ among blazar subclasses, exploring the apparent blazar sequence and evolution of the AGN \citep{fossati, ghisellini, meyer}.

VHE blazars can also be used to constrain the optical to near-infrared extragalactic background light (EBL) density and evolution, as well as the nature of cosmic rays.  The EBL encodes the integrated history of structure formation and the evolution of stars and galaxies in the Universe.  Understanding these characteristics requires detailed theoretical modeling of all the processes that contribute, e.g. structure formation and stellar evolution \citep{dominguez,gilmore09,primack,finkerazz,Franceschini08,hauser, stecker}.  According to some cosmic ray models, e.g. \cite{essey}, interactions of cosmic rays along the blazar line-of-sight can produce relatively hard VHE gamma-ray spectra as compared with the high-energy gamma-ray spectra, depending on the distance to the blazar.

The current catalog of extragalactic VHE objects largely contains relatively nearby AGN; only three have a confirmed redshift above $z=0.3$.  The highest redshift blazar detected at VHE thus far is the FSRQ 3C\,279, at a redshift of 0.536 \citep{3c279}.  The proximity of these VHE blazars is partly a result of gamma-ray absorption by the EBL.  VHE gamma rays that propagate through the intergalactic medium are absorbed by low energy EBL photons via pair production, $\gamma + \gamma \rightarrow e^+ + e^-$ \citep{nikishov,gould,steckerdejager}.  The absorption process deforms the VHE gamma-ray spectra emitted by extragalactic objects in a redshift-dependent manner and can be translated to upper limits on the local density of the far-infrared EBL \citep{steckeruls,dwek,aharonianNature,primack2011,orr}.

The search for new VHE blazars is complicated by the fact that many of these objects do not yet have known redshifts.  BL Lac objects, by definition, display very weak or no optical emission or absorption lines used for spectroscopic redshift measurements.  Due to the interaction of VHE photons with the EBL, non-detection of blazars with no redshift information can be either attributed to the object being too distant or having an intrinsically low luminosity in the VHE band.  This makes VHE discovery observations of blazars with no measured redshift a risky venture, although it has proven successful in the past, as in the discoveries of VHE emission from 3C\,66A, PKS\,1424+240, and RX\,J0648.7+1516 \citep{acciari3C66A,acciariPKS1424,furniss}.

The small field of view of imaging atmospheric Cherenkov telescopes (IACTs, $\le 5.0^{\circ}$) makes new source discovery from a large-scale sky survey difficult and therefore the hunt for VHE emitting objects has historically involved targeted observations of source candidates selected from surveys at lower frequencies, such as the selection of hard X-ray candidates presented in \cite{costamante}.  In this way, VHE blazar candidate selection has relied on experiments such as EGRET onboard the Compton Gamma Ray Observatory \citep{thompson}, \textit{Swift} \citep{gehrels} and ROSAT \citep{turriziani}.  The launch of \textit{Fermi} in June 2008 has enhanced VHE blazar discovery programs, leading to new blazar discoveries including RBS\,0413, RX\,J0648.7+1516, 1ES\,0033+595 and 1RXS J101015.9-311909 \citep{gunes,furniss,mariotti0033,abramowski}.

The \textit{Fermi} Large Area Telescope (LAT) observes the entire sky in the energy range from 20 MeV to $>$ 300 GeV every three hours and has better sensitivity than its predecessor EGRET.  Within the first 11 months of operation, the 1FGL catalog reported the detection of 1451 sources at a significance greater than 5$\sigma$ \citep{abdo1FGL}.  A majority of these sources are, or are expected to be, associated with AGN \citep{abdo1AGN}.  It is known that a large majority of these blazars has not yet been detected by VHE instruments, as can be seen by the current TeV catalog, which only contains 51 associated extragalactic objects.  However, the proximity of the \textit{Fermi}-LAT energy band to that of the IACTs makes the 1FGL catalog a good place to search for candidate VHE blazars.  We report on VERITAS observations above 100 GeV of six candidate blazars.  For the first time, a multiwavelength description of their SED including radio, UV, X-ray and gamma-ray frequencies is assembled and modeled using a SSC model.

\section{Target Selection}
The energy coverage of IACTs overlaps with that of \textit{Fermi}-LAT above 100 GeV and extends to tens of TeV. Sources that are most likely to be detected by both \textit{Fermi}-LAT and IACTs have high fluxes and hard photon indices in the \textit{Fermi}-LAT energy band. A selection process was established to identify the best VHE candidates within the first \textit{Fermi}-LAT catalog.  Sources at low Galactic latitude ($|$b$| < 10^{\circ}$) were excluded, with the intent of removing the majority of Galactic sources from the selection.  Additionally, sources with low integrated flux above 100 MeV (F$_{100 MeV} < 2 \times 10^{-9} $ph cm$^{-2}$s$^{-1}$), with soft photon indices ($\Gamma >$ 2.0), or with a low number of associated photons ($N_{pred} < 20$), were excluded from the selection.  Just over 200 sources passed this initial set of cuts.

The remaining sources were re-analyzed using the same data set as that used for the 1FGL catalog to search for the presence of curvature in their spectra. A log-parabola parameterization for the spectra was chosen as the alternative to the power-law model (null hypothesis). In addition, the data were analyzed only using data above 1 GeV to confirm that the results from the power-law fit found for the entire \textit{Fermi}-LAT energy band agreed well with the power-law fit above 1 GeV.  Sources showing significant spectral curvature or softening of the spectrum above 1 GeV with an improvement in likelihood value corresponding to 3$\sigma$ were excluded from the final selection as these sources are not expected to exhibit bright TeV flux levels. For sources with spectra best fit by a power law that also matched the power-law fit above 1 GeV, the \textit{Fermi}-LAT spectrum was extrapolated up to the VHE regime (150 GeV - 1 TeV). When a reliable redshift measurement was available, the EBL model of \cite{Franceschini08} was used to estimate the extrapolated VHE gamma-ray flux. Otherwise an assumed redshift of z = 0.2 was used, a conservative value considering that most known TeV sources have redshifts less than $z=0.2$. Finally, the sources were ranked based on their extrapolated flux in the VHE regime. Six VHE candidates, all BL Lac objects, were selected for observations with the VERITAS telescope between September 2009 and June 2010 for 10 hours each, corresponding to the VERITAS 3\% Crab flux sensitivity exposure timescale.

Table 1 shows the VHE extrapolated integral flux from the 1FGL catalog power-law fits used for selection of the six candidate VHE emitting blazars.  These relatively high integral flux values above 150 GeV, shown in percentages of the Crab Nebula flux\footnote{Flux calculated according to the curved power law presented in \cite{albertCrab}}, are shown in comparison to the updated power-law fits from 29 months of \textit{Fermi}-LAT data for the steady sources, and data from the VERITAS coincidental window for variable sources. The analysis of the extended \textit{Fermi}-LAT dataset (2008 August 4 to 2011 January 4) is detailed in section 4.2.  These updated extrapolations show much lower expected integral flux values for each of the six candidates, reflecting the fact that as more LAT data were collected, better high-energy statistics showed the candidates to be softer and/or dimmer than found with the elevent month data set used in the 1FGL.  Additionally, we compare these extrapolated values to the upper limits derived from the VERITAS observations, where the analysis leading to these upper limits is detailed in Section 4.1.

\section{The Targets}
\textbf{RGB\,J0136+391} (1FGL\,J0136.5+3905) was discovered in the third Bologna sky survey of 408 MHz radio objects \citep{ficarra}.  It was later detected as a X-ray bright active galaxy in the Northern ROSAT all-sky survey \citep{brinkmann}, and identified spectroscopically as a BL Lac object, with an IBL sub-classification from \cite{laurent} and \cite{laurentB}, respectively.  Based on the optical and X-ray spectral properties, this blazar was proposed as a VHE candidate blazar by \cite{costamante}, under the assumption that the redshift was less than 0.2.  Bright gamma-ray emission above 1 GeV was detected from this source by \textit{Fermi} after three months of operation.  Only an upper limit below 1 GeV was reported \citep{abdo09}.  This blazar also showed constant emission in the first eleven months of LAT operation \citep{abdo1FGL}.  This blazar remains without a known redshift, with no previous spectroscopic redshift measurements found in the available literature.  We attempted a spectroscopic redshift measurement on 2009 September 17 (MJD 55091) using the Keck Low Resolution Imaging Spectrometer (LRIS) instrument, but measured only a featureless power-law spectrum that is characteristic of BL Lac objects (see Figure 1 upper-left panel).  Detailed spectroscopic analysis of this data can be found in \cite{kaplan}. 

\textbf{RGB\,J0316+090} (1FGL\,J0316.1+0904) was first detected by the NRAO Green Bank 91 meter radio telescope \citep{becker}. It was later optically identified as a BL Lac object \citep{fischer} and sub-classified as an IBL by \cite{laurentB}.  No spectroscopic redshift measurements had been made prior to this work.  We attempted a spectroscopic redshift measurement on 2011 March 5 using the Keck Echelette Spectrograph and Imager (ESI) instrument, resulting in a featureless power-law spectrum except for two unidentified absorption features (see Figure 1 upper-right panel).  The detailed spectroscopic analysis of the Keck ESI data can be found in \cite{kaplan}.

\textbf{RGB\,J0909+231} (1FGL\,J0909.2+2310) was first detected by the NRAO Green Bank radio telescope \citep{becker} and was later classified as a radio-loud active galaxy \citep{brinkmann}.  The BL Lac optical counterpart was identified nearly a decade later \citep{mickaelian}.  The redshift reported by the NASA Extragalactic Database (NED)\footnote{\tt http://nedwww.ipac.caltech.edu/} could not be found within the corresponding reference ($z=0.231$, \cite{brinkmannA}) and was therefore taken as unknown.  Inspection of publicly available Sloan Digital Sky Survey (SDSS) data revealed two Mg II absorption lines in the otherwise featureless optical spectrum.  Assuming that these lines could be intrinsic, or due to the absorption by an intervening cloud, a lower limit on the redshift of $z\ge0.43$ is derived (see Figure 1 middle-left panel).   This is the only source out of the six selected which has a neighboring \textit{Fermi}-LAT detected blazar within the VERITAS 3$^{\circ}$ field of view. 2FGL\,J0910.9+2246, associated with TXS\,0907+230, is located 0.61$^{\circ}$ away from RGB\,J0909+231.  The high redshift blazar TXS\,0907+230 ($z=2.66$ according to \cite{healey}) shows no signal in the 10-100 GeV band, with an upper limit of 9.6$\times10^{-11}$photons cm$^{-2}$s$^{-1}$ reported in the 2FGL catalog \citep{2fgl}.  An exclusion region of radius 0.3$^{\circ}$ centered on the blazar was nevertheless used in the VERITAS analysis to avoid possible contamination.

\textbf{RGB\,J1058+564} (1FGL\,J1058.6+5628) was first detected in the 6C radio survey \citep{hales}.  It was identified as a BL Lac object during the association of the ROSAT all-sky survey with the Hamburg Quasar Survey \citep{nass}.  \cite{bondi} classified the object as an IBL based on the optical and X-ray characteristics.  This blazar was detected within the first three months of \textit{Fermi}-LAT operation \citep{abdo09} with a broadband SED shown in \cite{abdoSED} which does not include any VHE information.  The blazar also shows a high level of flux variability (probability of variability: 79\%) in the high-energy gamma-ray band, as shown in \cite{abdo1FGL} and is the only one out of the six targets in this paper that has a redshift measurement.  The absorption lines corresponding to the redshift 0.143 can be seen in the SDSS spectrum shown in Figure 1 (middle-right panel).  

\textbf{RGB\,J1243+364} (1FGL\,J1243.1+3627) was first reported in the B2 catalog of radio sources \citep{colla}.  This target was also determined to be a radio-loud active galaxy by ROSAT \citep{brinkmann}, and specifically classified as a BL Lac in \cite{appenzeller}.  NED cited the SDSS data for a spectroscopic redshift of $z=1.065$.  Inspection of this publicly available SDSS data revealed no lines suggesting this redshift but instead revealed Mg II absorption lines that translate to a lower limit of $z\ge0.485$ (see Figure 1 lower-left panel).  A recent photometric redshift of z$=0.5^{+0.14}_{-0.12}$ from \cite{meisner} is in agreement with the SDSS lower limit. 

\textbf{RX\,J1436.9+5639} (1FGL\,J1437.0+5640) was detected in the ROSAT all-sky survey and identified as a BL Lac object by \cite{nass}.  This HBL, as classified by \cite{nieppola} based on the frequency of the synchrotron peak, remains without a redshift, although the redshift of the galaxy cluster within the same region of the sky is known to be z$=0.15$ \citep{bauer}.   Inspection of the publicly available SDSS data shows a featureless spectrum, shown in Figure 1.

\section{Multiwavelength Observations and Analysis}
\subsection{VERITAS}
The VERITAS observatory is an array of four 12-meter diameter IACTs, located in southern Arizona.  VERITAS is sensitive to photons between 100 GeV and several tens of TeV with an energy resolution of better than 20\%.  The instrument has a 5$\sigma$ point source sensitivity of 1\% of the Crab Nebula flux in less than 30 hours with an angular resolution of less than 0.1$^{\circ}$ for a Crab-like source with a spectral index of 2.5.  See \cite{weekes} and \cite{holder} for a detailed overview of the instrument.

The VERITAS observations of the six VHE candidate blazars were completed between September 2009 and June 2010 (MJD 55122--55383).  These observations were taken in \textit{wobble} mode, with an offset of $0.5^{\circ}$ from the source position in each of four cardinal directions to allow simultaneous background measurement, as explained in \cite{fomin} and \cite{berge}.   The radio location of the counterparts as specified by NED were used for source position.

Air shower events initiated by gamma and cosmic-rays are reconstructed following the procedure outlined in \cite{acciari2008}. The recorded shower images are parameterized by their principal moments, giving an efficient method for suppression of the far more abundant cosmic-ray background.  Any events with a total charge less than 50 photoelectrons are removed from the analysis.  A set of cuts is then applied to the parameters in order to reject background events (see details of this method in \cite{krawczynski}).  These parameters reject cosmic-ray like events having the \textit{mean scaled width} and \textit{length} of the event camera image smaller than 1.1 or 1.4, respectively.   Additionally, the reconstructed altitude of the maximum Cherenkov emission is required to be higher than 8 km above ground level.   Gamma-ray like events are extracted from a signal region with a radius of 0.14$^{\circ}$, centered at the coordinates of the candidate source.

The quality-selected livetime collected for each target ranges from 4.4 to 14.2 hours and results in no detections, with significances ranging from -1.1 to 0.9$\sigma$, calculated with Equation (17) of \cite{lima}.  The VERITAS observations and analysis results are detailed in Table 2.  Integral upper limits at 99\% confidence level are calculated using the \cite{rolke} method, assuming a photon index of $\Gamma=3.0$ for the differential power-law spectrum $dN/dE = N(E/E_o)^{-\Gamma}$.  This index was assumed as a moderately softened index as compared to the \textit{Fermi}-LAT index range (1.74$-$2.9), expected due to absorption by the EBL and possible intrinsic turnover.  Additionally, this index value is representative of a typical TeV blazar.  Without detection, the real index in the VHE band remains unknown and any integral upper limit derived is dependent on the index assumed.  The differential upper limit is quoted at the decorrelation energy, the energy where the calculated flux has minimal dependence on the index.  Changing the spectral index by $\pm$0.5 changes the differential upper limits by less than 10\%.

The results are independently reproduced with two different analysis packages, as described in \cite{cogan} and \cite{daniel}.  The upper limits range from 1 to 3\% of the integral Crab Nebula flux above the threshold energy.  The energy threshold for each observation is defined as the energy at which the differential rate of reconstructed gamma rays from the postulated source reaches its maximum and is accurate to within the 20\% energy resolution of the instrument, a value that is dependent on the observation zenith angle and sky brightness.

\subsection{\textit{Fermi} LAT}
The \textit{Fermi}-LAT is a pair-conversion telescope sensitive from 20 MeV to $>$ 300 GeV, which operates in survey mode. Further details about the characteristics and performance of the LAT Instrument on the \Fermi Gamma-ray Space Telescope can be found in \citet{atwood}. Presented here is the analysis of the \FermiLAT data for the six candidates described in
Section 3. Although the targets were selected based on only 11 months of data, more data was available after the completion of VERITAS observations and this larger data set is used for the modeling.  More specifically, the LAT data from the time period of 2008 August 4 to 2011 January 4 (MJD 54682.7-55565.0) were used for the modeling analysis.   Except for variable sources, the analysis
procedure was identical for each of the sources and proceeded as follows.

For each candidate, events were extracted from a region of interest (ROI) of 10$^{\circ}$ radius centered on the target coordinates. Events from the ``diffuse class'' with zenith angle $< 100^{\circ}$ and energy between 300 MeV and 300 GeV were selected. Only data taken during periods when the rocking angle of the satellite was $< 52^{\circ}$ were used to reduce contamination from the Earth limb gamma rays, which are produced by cosmic rays interacting with the upper atmosphere.  The significance and spectral parameters were calculated using an unbinned maximum-likelihood method implemented in the LAT Science Tool \texttt{gtlike}\footnote{\texttt{ScienceTools v9r20p0} with the post-launch instrument response function (IRF) \texttt{P6\_V11\_DIFFUSE}.} \citep{cash,mattox}.  A background model was constructed including nearby gamma-ray sources and diffuse emission.  All sources within 12$^{\circ}$ of the central source in the second \textit{Fermi}-LAT catalog  (2FGL, \cite{2fgl}) were included in the model.  The spectra of known pulsars were modeled by a power laws with exponential cutoffs.  As in the 2FGL catalog, a log-parabola function was used for sources with significant spectral curvature.  Otherwise, spectra were described as a power law. The spectral parameters of the sources in the ROI were left free during the fitting procedure.  Sources outside the ROI, but within the 12$^{\circ}$ range had their spectral parameters fixed to the 2FGL catalog values. The Galactic diffuse emission and an isotropic component, which is the sum of the extragalactic diffuse gamma-ray emission and the residual charged particle background, were modeled using the recommended files\footnote{The files used were \texttt{gll\_iem\_v02\_P6\_V11\_DIFFUSE.fit} for the Galactic diffuse and \texttt{isotropic\_iem\_v02\_P6\_V11\_DIFFUSE.txt} for the isotropic diffuse component available at \tt http://fermi.gsfc.nasa.gov/ssc/data/p6v11/access/lat/BackgroundModels.html}.

The LAT data were first analyzed to calculate the time-averaged gamma-ray flux and spectral parameters of each candidate. A second analysis was then performed to study the impact of the Sun, a bright gamma-ray source, on the flux of candidates located near the plane of the ecliptic (RGB\,J0316+090 and RGB\,J0909+231). Removing time intervals when the Sun was in the ROI of each candidate had a negligible effect on the analysis results.

Spectral points and a light curve were calculated for each candidate, and a temporal analysis was performed to search for flux variability. The timescale of this analysis was adjusted based on the specific candidate flux levels. The flux in each energy or time bin was determined with the spectral indices of all sources fixed to the best-fit values over the full energy and time interval. For an energy or time bin with a test statistic (TS; see \cite{mattox}) less than 9 or fewer than 3 predicted photons ($N_{pred}$), a 95\% confidence level upper limit was calculated.

The light curves were analyzed to search for flux variability with a likelihood method assuming a constant flux for the null hypothesis, following the same procedure as used in the 2FGL catalog. Only two sources showed significant evidence of flux variability: RGB\,J0316+090 and RGB\,J1058+564. For these two sources, a refined analysis was done, selecting time periods contemporaneous with the VERITAS observation windows.   The duration of this contemporaneous period was chosen such that a significant detection ($> 5\sigma$) could be attained, resulting in slightly extended windows of MJD 55055-55145 for RGB\,J0316+090 and MJD 55160-55185 for RGB\,J1058+564 with respect to the VERITAS observation window.  For these candidates, a butterfly corresponding to the $1\sigma$ confidence interval was used to represent the spectral information (Figure 3).

The derived spectral indices for the differential power laws obtained for the candidates are relatively hard, ranging from 1.74 to 2.09, as compared to the 2FGL average spectral index of 2.21 $\pm$ 0.01. The integral fluxes above 300 MeV range from 1.55 to 12.4$\times10^{-9}$ph cm$^{-2}$s$^{-1}$, indicating that these six blazars are bright in the high-energy band.  The detailed \textit{Fermi}-LAT results for each of these hard-spectrum, bright BL Lacs are summarized in Table 3.

\subsection{\textit{Swift} XRT}
The X-ray Telescope (XRT) onboard the \textit{Swift} satellite \citep{gehrels} is a focusing X-ray telescope sensitive to photons with energy between 0.2 and 10 keV. The data used for the broadband SED modeling were analyzed as described in \cite{burrows} with the \texttt{HEASoft} package Version 6.9 and \texttt{XSPEC}\footnote{\tt http://heasarc.nasa.gov/docs/software/lheasoft/xanadu/xspec/XspecManual.pdf} Version 12.6.0.  All data were taken in photon counting mode and pile up effects are accounted for when count rates exceeded 0.5 counts per second through the use of an annular source region, with a 1-2 pixel inner radius and a 20 pixel outer radius. Each observation is binned and fit with an absorbed power law between 0.3 and 10 keV, with the neutral hydrogen density taken from the Leiden/Argentine/Bonn survey of Galactic HI \citep{kalberla}. 

X-ray variability is commonplace for both VHE detected and non-detected blazars.  If more than one exposure exists for an object and no variability is detected, the de-absorbed power-law fit of the combined data set is used to constrain the SED modeling.  If variability is observed between multiple exposures, results from these separate exposures are shown independently on the SED plot, as is the case for RGB\,J0136+391 and RGB\,J1058+564.  These blazars show flux variability factors of $\sim$2 and 3 between exposures, respectively. The multiple exposures taken on RGB\,J0909+231 do not provide sufficient statistics for application of an absorbed power-law model and are therefore summed before fitting.  Using Cash analysis \citep{cash} did not improve the fitting of the single low-statistics spectra.  The summed exposure fit result is shown on the SED.

Each of the absorbed power-law fits applied to the XRT data resulted in photon indices greater than 2, with 2-10 keV integral flux levels between 0.3 and 21$\times10^{-12}$ergs cm$^{-2}$s$^{-1}$.  The index values suggest that the synchrotron component peaks below keV energies, characteristic of ISP blazars.  The analysis results for each observation are summarized in Table 4.

\subsection{\textit{Swift} UVOT}
The \textit{Swift}-XRT observations were supplemented with simultaneous UVOT exposures taken in the V, B, U, UVW1, UVM2, and UVW2 bands \citep{poole}.  The UVOT photometry is performed using the \texttt{HEASoft} program \textit{uvotsource}.  The circular source region has a $5\arcsec$ radius and the background region consists of several 15$\arcsec$ radii circles of nearby empty sky. The results are reddening corrected using E(B-V) coefficients \citep{schlegel}.  The Galactic extinction coefficients are applied according to \cite{fitzpatrick}.  The uncertainty in the reddening E(B-V) is the largest source of error, especially in the UV bands for blazars that have a large value of E(B-V). If more than one exposure exists in a specific band for an object, the data from the observation closest to the VERITAS exposure are used, although no significant variability is seen across any band for any blazar.  A summary of the UVOT analysis results is presented in Table 5.

\section{Broadband SSC Modeling}
Leptonic models for blazar jet emission attribute the higher energy
peak in the SED to the inverse-Compton scattering of lower
energy photons off a population of non-thermal, relativistic
electrons.  These same electrons are responsible for the lower-energy
synchrotron emission that makes up the first peak.  The target photon
field involved in the Compton upscattering can either be the synchrotron
photons themselves, as is the case in SSC models, or a photon field external to the jet in the case for EC models.  

The previously described multiwavelength data are matched with archival radio data
collected from NED.  These data are
used to test a steady-state leptonic jet model for
the broad-band continuum emission from the blazars.  Although it has been found that ISP BL Lacs are sometimes better represented by external-Compton models, e.g. \cite{acciariWComae}, taking into account the lack of redshift information and lack of constraints from the broadband data we prefer not to apply an EC model, which would introduce additional free parameters, compared to the SSC model applied here.  The model-predicted flux reflects the absorption of VHE gamma rays by the EBL according to the redshift information summarized in Table 1 using the model from \cite{gilmore09}.  The level of TeV absorption resulting from this model is consistent with the absorption derived from the \cite{finkerazz} and \cite{Franceschini08} models.

The SSC model applied to the multiwavelength data is the equilibrium version of the
\cite{bc02} model, as described in \cite{acciariWComae}.  In this model,
the emission originates from a spherical blob of relativistic
electrons with radius $R$.  This blob is moving down the jet with a 
Lorentz factor $\Gamma$, which corresponds to a speed of
$\beta_{\Gamma} c$.  The jet axis is aligned toward the line of sight with an angle $\theta_{\rm obs}$, which results
in Doppler boosting with a Doppler factor $D = (\Gamma [1-\beta_{\Gamma}
\cos\theta_{\rm obs}])^{-1}$.  In order to minimize the number of
free parameters, we assume that $\theta_{obs} = 1/\Gamma$,
often referred to as the critical angle, for which $\Gamma = D$.

Within the model, nonthermal electrons are injected and accelerated
into a power-law distribution $Q(\gamma) = Q_0
\gamma^{-q}$ between the low- and high-energy cut-offs,
$\gamma_{min}$ and $\gamma_{max}$.  The radiation mechanisms considered lead to an
equilibrium between particle injection, radiative cooling and
particle escape.  This particle escape is characterized with an escape
efficiency factor $\eta$, such that the time scale of escape $t_{\rm
esc} = \eta \, R/c$.  This results in a particle distribution
which streams along the jet with a power $L_e$.  Synchrotron emission
results from the presence of a tangled magnetic field $B$, with a
Poynting flux luminosity of $L_B$.

The two parameters $L_e$ and $L_B$ allow the calculation of the equipartition parameter
$\epsilon_{Be} \equiv L_B/L_e$.  This equipartition parameter is used as an estimator of the feasibility of the model, where models which result in $\epsilon_{Be} \sim 1$ are preferred.  If the particle energy density greatly dominates over the magnetic field energy density, namely a particle dominated scenario, then the magnetic field cannot serve to collimate the jet.  Following this design, acceptable parameters should result in at least partition conditions with $L_B\ge L_e$. 

The broadband SED for each blazar can be seen in Figure 3, with the SSC model parameters for each representation summarized in Table 6.  For each blazar, the archival radio data are taken as upper limits as these measurements are believed to contain a large amount of radiation produced in the radio lobes in addition to the synchrotron emission from the jet.  The modeling for each of the six blazars shows synchrotron peak locations $\sim10^{15}$ Hz, characteristic of borderline ISP/HSP blazars. 

\textbf{RGB\,J0136+391:} This blazar is modeled using the lower of the two variable X-ray states for three different assumed redshifts ($z=0.2, 0.3$ and $0.4$).  The variability timescale of three months suggested by the factor-of-two variability observed between XRT exposures is not short enough to provide a constraint on the size of the model emission region.  Under the assumption that the gamma-ray emission remained steady during the \textit{Fermi} and VERITAS observations, the hard LAT spectrum and low VERITAS upper limit derived for this blazar suggest a steepening of the gamma-ray spectrum at $E\ge100$ GeV, which could be caused by the EBL absorption if a redshift $z\ge0.4$ is assumed.  Alternatively, this apparent break could originate from uncorrelated variability in the high and very-high energy bands.  More specifically, a low flux state in the VHE band during VERITAS observations could provide a redshift-independent explanation of the apparent spectral softening.  Figure 3 shows the model predictions, corrected by EBL absorption, for each of the redshift values.  Only the model at $z=0.4$ is compatible with the VERITAS upper limit, which also results in a framework with balanced radiation and particle energy.  

\textbf{RGB\,J0316+090:}
Due to the variability detected in the high-energy band, this blazar is modeled with \textit{Fermi}-LAT data which spans the complete time period sampled by shorter VERITAS observations.  The UVOT errors resulting from the E(B-V) reddening correction for this blazar are so large that all exposures except the V band are unconstraining to the SED modeling and therefore not shown.  Application of the SSC model for an assumed redshift of $z=0.2$ results in a particle dominated framework consistent with the VERITAS upper limit.  

\textbf{RGB\,J0909+231:}
The SSC model parameters used to describe the broadband emission of this blazar are determined for a redshift of $z=0.5$, based on the lower limit derived from Mg II absorption lines found in the SDSS data.  The \textit{Swift} X-ray data from three exposures have been summed in order to provide sufficient statistics for absorbed power-law fitting.  The model results in a particle-dominated scenario with a slight discrepancy between the model and the \textit{Fermi} upper limit in the $1-3$ GeV energy bin.  Moving the Compton peak accomodates this upper limit only results in an even less favored, particle-dominated emission state.  

\textbf{RGB\,J1058+564:}
Due to the variability observed in the LAT band, the SED modeling for this blazar is done with LAT data from the time period coincident with the VERITAS observation window and for the lower of two X-ray states observed.  The variability timescale suggested by the variability factor of three observed between XRT exposures is not short enough to provide any constraint on the size of the emitting region.  Both X-ray states are shown on the SED for reference.  The SSC model shows agreement with the broadband data, although results in a disfavored particle dominated scenario. 

\textbf{RGB\,J1243+364:}
The modeling for this blazar is completed for a redshift of $z=0.5$, based on the new lower limit found from Mg II absorption lines in the public SDSS data.  The model agrees with the broadband data and permits parameters at equipartition.  

\textbf{RX\,J1436.9+5639:}
The modeling for the broadband data of this blazar allows marginal agreement with the \textit{Fermi} data when completed for the redshift of the spatially coincident galactic supercluster, more specifically $z=0.15$.  The resulting model parameters are far below equipartition, suggesting a particle dominated jet.

\section{Discussion}
Six promising TeV blazar candidates were selected from the 1FGL catalog for observations with VERITAS.  These observations resulted in no VHE detections. Initial selection of these candidates from 1FGL power-law fit extrapolation suggested integral flux levels between 1 and 12\% of the Crab Nebula flux above 150 GeV after accounting for absorption by the EBL.  With additional \textit{Fermi}-LAT data and more information about the blazar redshifts, the expected fluxes were updated to levels between 0.3\% and 3.5\% of the Crab Nebula flux above 150 GeV.  The VERITAS exposure times were allocated based on the initial extrapolated values, resulting in only one upper limit below the updated VHE extrapolation, namely the upper limit for RGB\,J0136+391.  The non-detection of this blazar suggests spectral steepening of the high-energy spectrum that can either be explained by intrinsic spectral curvature, redshift-dependent EBL absorption or uncorrelated variability.

Multiband observations are presented and allow for the construction and SSC modeling of the radio through TeV broadband SEDs.  The model applied to these blazars is sufficient to represent the broadband data for each of the six, with model parameters roughly comparable to those found for other VHE-detected ISPs using the same model \citep{abdo3C66A,acciariWComae}.  The magnetic fields obtained in the modeling are generally low, resulting in disfavored, particle-dominated jets.  This condition could be relaxed by including an external photon field for inverse-Compton scattering, allowing solutions closer to equipartition, as was done in \cite{abdo3C66A} and \cite{acciariWComae}.  However, this scenario has not been explicitly tested.  The quality of the data sampling of the inverse-Compton peak and the fact that the redshifts are not well determined for the majority of the objects do not provide sufficient constraints for a model with the extra degrees of freedom associated with adding an external source of seed photons.  Similar parameters were also found for the borderline ISP/HSP TeV detected blazar PKS\,1424+240, with the exception of the spectral index for the injected electron distribution, which was found to be very soft ($q$ = 5.1, \cite{acciariPKS1424}), while the typical values for this model fall between $q=$2.3 and 2.7.  

Each of these blazars shows a synchrotron peak frequency characteristic of the ISP/HSP divide, namely $\nu_{synch}\sim10^{15}$.  Additionally, each of these BL Lacs exhibits a similar flux level within the high-energy gamma-ray band, showing comparable L$_{synch}$ and L$_{IC}$.  This commonality of sub-type and peak luminosities is likely a bias introduced to the selection process through the sensitivity of the \textit{Fermi} LAT instrument being greater in the 300 MeV - 100 GeV band as compared to the $\ge$100 GeV band.   Nearby blazars releasing a large fraction of power above 100 GeV are known to be good candidates for detection in the VHE regime.  This type of power emission is characteristic of HSP BL Lacs, the most commonly VHE-detected type of blazar, but the least frequently detected BL Lac in the 1FGL and 2FGL catalogs. 

Although the selection of these six blazars from high-energy \textit{Fermi}-LAT data did not lead to any new TeV blazar discoveries, the exercise has been very illuminating.  We are reminded that when selecting candidates for observation with TeV instruments, accurate redshift information is crucial.  Without this information it is difficult to decipher if the lack of TeV emission is due to the intrinsic emission mechanism or the absorption of gamma rays by the EBL.  The redshift lower limits that were found in the SDSS for RGB\,J0909+231 and RGB\,J1243+364 indicate distances where future TeV detection from a deeper exposure is unlikely.  Continued TeV observations of RGB\,J0136+361, RGB\,J0316+090 and RX\,J1436.9+5639, the three sources that remain without solid redshift information, could provide insight into the distance to these sources, while further observations of RBG\,J1058+564 can be directly applied to further investigate the emission mechanism at work within the blazar jet. 

\acknowledgments
This research is supported by grants from the U.S. Department of Energy Office of Science, the U.S. National Science Foundation and the Smithsonian Institution, by NSERC in Canada, by Science Foundation Ireland (SFI 10/RFP/AST2748) and by STFC in the U.K. We acknowledge the excellent work of the technical support staff at the Fred Lawrence Whipple Observatory and at the collaborating institutions in the construction and operation of the instrument.

The \textit{Fermi} LAT Collaboration acknowledges generous support
from a number of agencies and institutes that have supported the
development and the operation of the LAT as well as scientific data analysis.
These include the National Aeronautics and Space Administration and the
Department of Energy in the United States, the Commissariat \`a l'Energie Atomique
and the Centre National de la Recherche Scientifique / Institut National de Physique
Nucl\'eaire et de Physique des Particules in France, the Agenzia Spaziale Italiana
and the Istituto Nazionale di Fisica Nucleare in Italy, the Ministry of Education,
Culture, Sports, Science and Technology (MEXT), High Energy Accelerator Research
Organization (KEK) and Japan Aerospace Exploration Agency (JAXA) in Japan, and
the K.~A.~Wallenberg Foundation, the Swedish Research Council and the
Swedish National Space Board in Sweden.

Additional support for science analysis during the operations phase is
acknowledged from the Istituto Nazionale di Astrofisica in Italy and the Centre National d'\'Etudes Spatiales in France.

J.X.P. acknowledges
funding through an NSF CAREER grant (AST--0548180).

Funding for the SDSS and SDSS-II has been provided by the Alfred P. Sloan Foundation, the Participating Institutions, the National Science Foundation, the U.S. Department of Energy, the National Aeronautics and Space Administration, the Japanese Monbukagakusho, the Max Planck Society, and the Higher Education Funding Council for England. The SDSS Web Site is \url{http://www.sdss.org/}.

This research has made use of the NASA/IPAC Extragalactic Database (NED) which is operated by the Jet Propulsion Laboratory, California Institute of Technology, under contract with the National Aeronautics and Space Administration.

{\it Facilities:} \facility{VERITAS}, \facility{Fermi-LAT}, \facility{Swift XRT and UVOT}, \facility{SDSS}, \facility{Keck}, \facility{Lick}

\clearpage


\begin{figure}
\epsscale{0.69}
\plotone{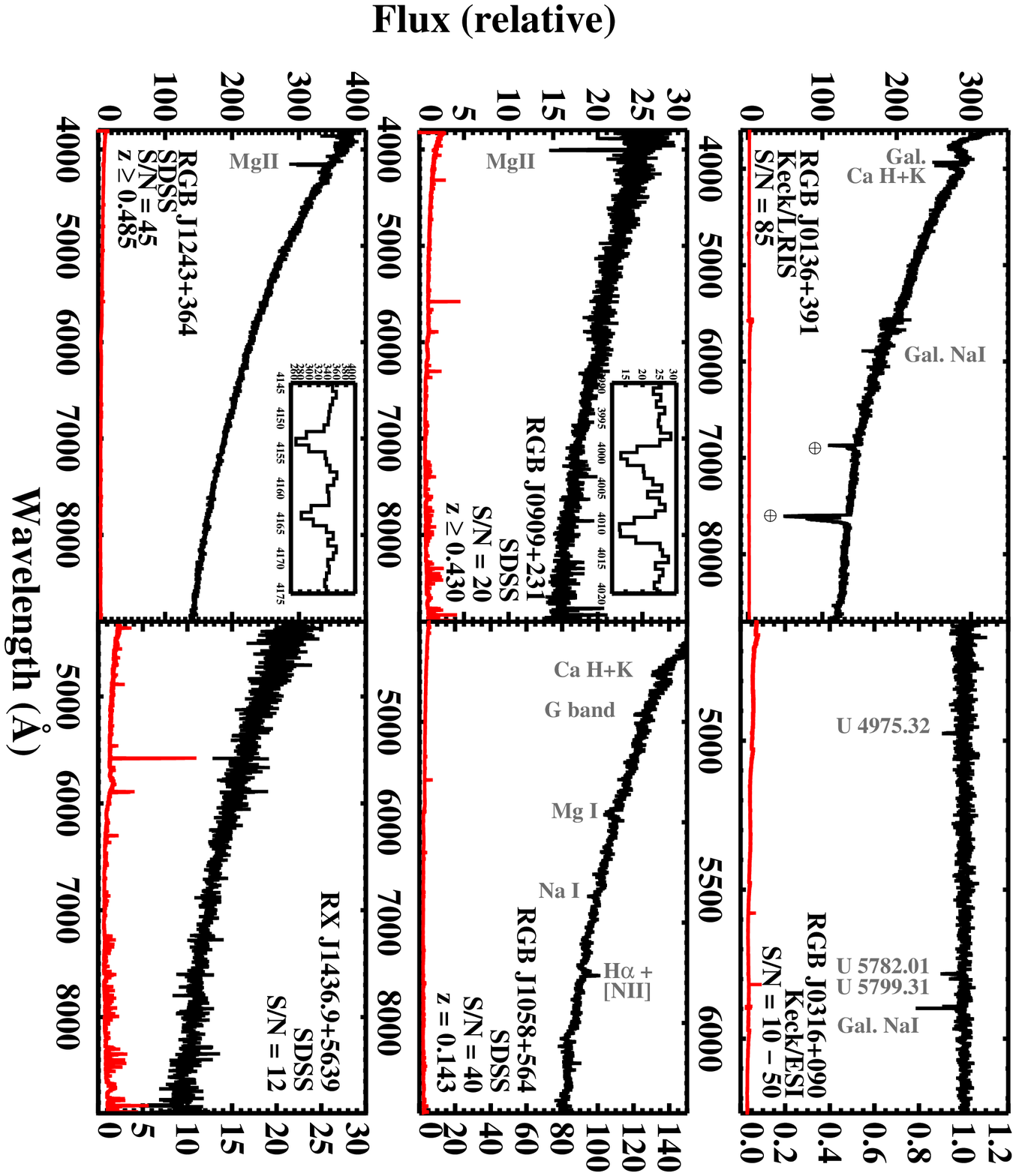}
\caption{Optical spectra for the six BL Lacs selected from the 1FGL catalog and observed with VERITAS.  The black shows the object spectrum, while the red shows the instrumental noise.  Only one BL Lac had a confirmed redshift upon selection (RGB\,J1058+564, z=0.143), confirmed with the SDSS spectrum shown in the middle-right panel.  Redshift lower limits for RGB\,J0909+231 ($z\ge0.4305$; middle left) and RGB\,J1243+364 ($z\ge0.485$; lower left) are found from Mg II absorption lines in the SDSS spectra.  A featureless SDSS spectrum is found for RX\,J1436.9+5639.  A redshift measurement attempt for RGB\,J0136+391 using the Keck LRIS instrument shows a featureless power-law spectrum (upper left).  The normalized ESI spectrum of RGB\,J0316+090 exhibits three unidentified absorption features (upper right).  The spectral analysis of the Keck LRIS and ESI spectra are detailed in \cite{kaplan}. \label{fig1}}
\end{figure}
\clearpage

\begin{figure}
\epsscale{1.0}
\plotone{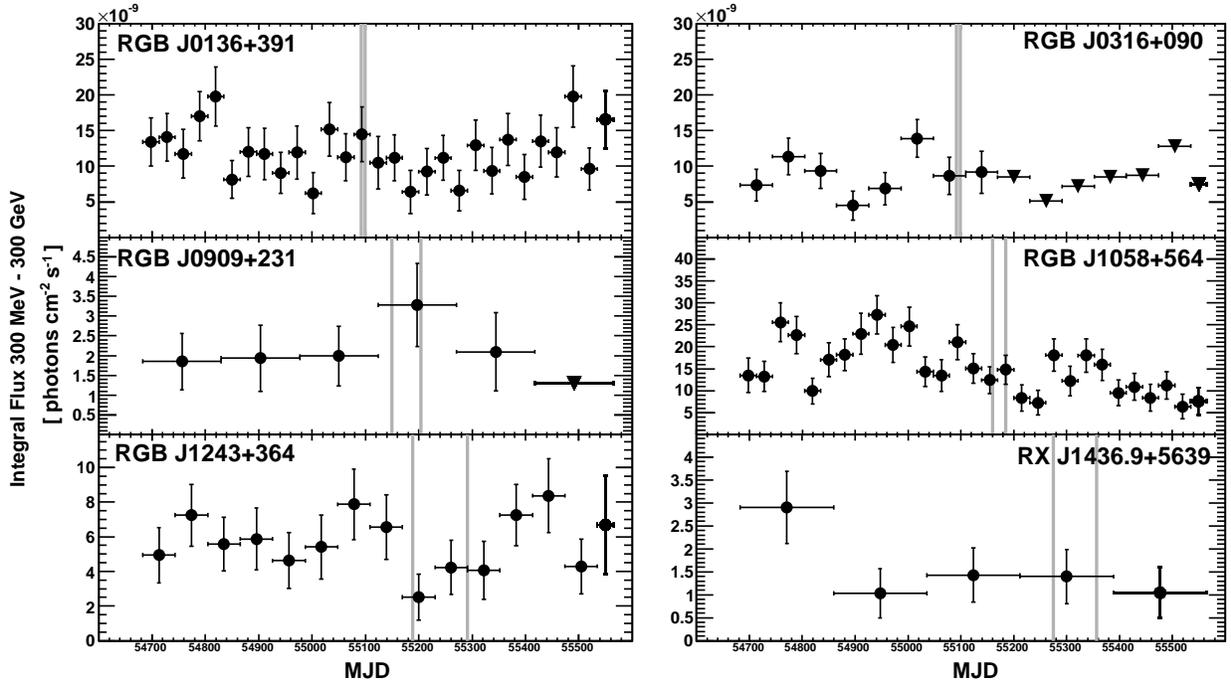}
\caption{\textit{Fermi}-LAT light curves, with units of 10$^{-9}$ ph cm$^{-2}$s$^{-1}$, are shown for the six candidate VHE-emitting BL Lacs for 29 months of LAT data (MJD 54682-55565; 2008 August 4-2011 January 4).  The beginning and end of the VERITAS observations are denoted by vertical grey lines in each panel.  The short VERITAS observation periods for RGB\,J0136+364 and RGB\,J0316+090 can be seen, representing only 7 days each. Upper limits at 95\% confidence level are shown for bins resulting in TS of less than 9, denoted by downward pointing black triangles.  Only RGB\,J0316+090 and RGB\,J1058+564 show any significant variability.  For these two sources, LAT data only for the time periods within the window of VERITAS observations are used to constrain the modeling. \label{fig2}}
\end{figure}
\clearpage

\begin{figure}
\epsscale{1.0}
\plotone{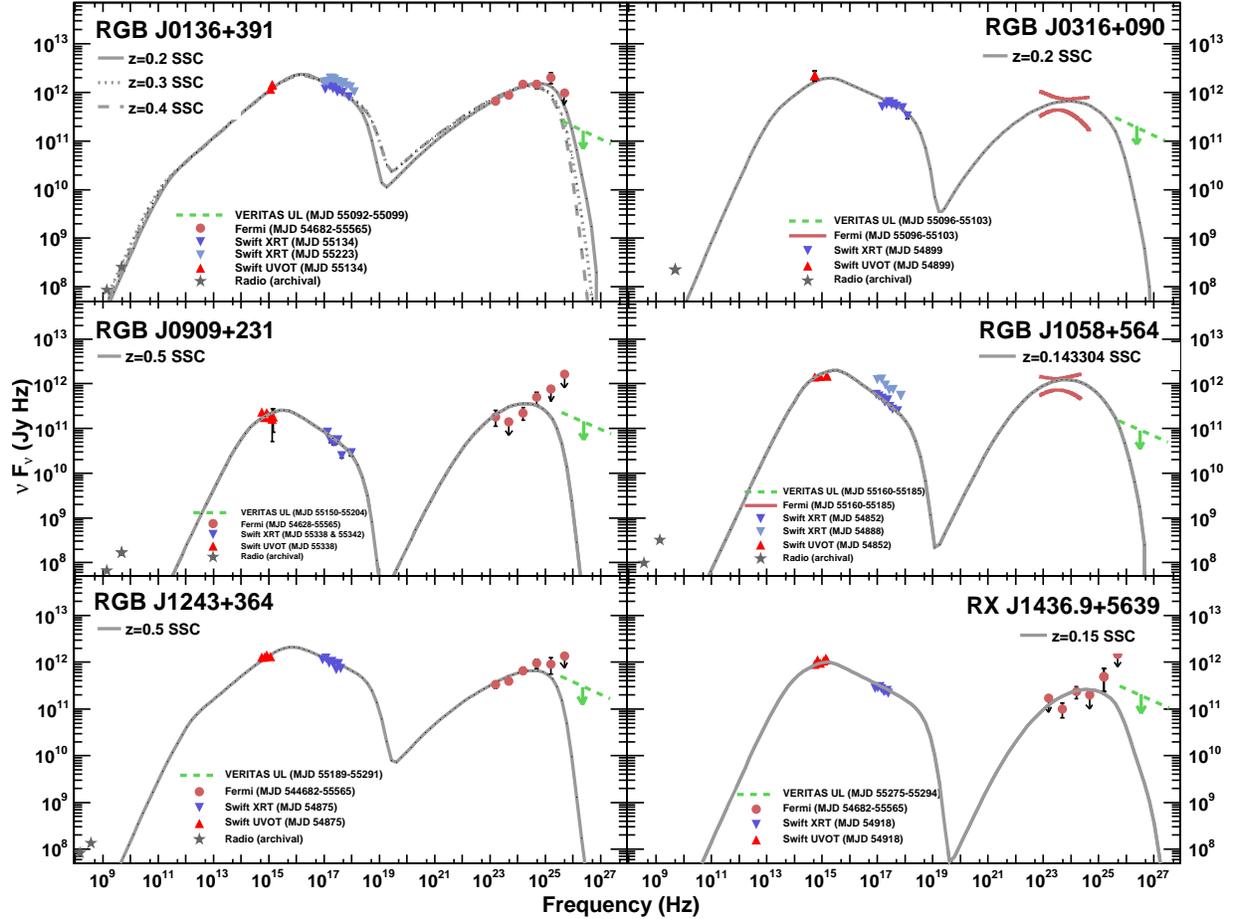}
\caption{Non-contemporaneous broadband SED data for each BL Lac shown with corresponding SSC modeling using the model of \cite{bc02}.  The modeling results, corrected for the EBL absorption, are shown with grey lines.  The archival radio data points are taken from NED and used only as upper limits, accounting for the fact that much of the radio emission may result from diffuse synchrotron emission in the radio lobes of the jet.  See the text for a more detailed description of the model parameters, with values listed in Table 5.}
\end{figure}
\clearpage


\begin{deluxetable}{ccccccccc}
\tabletypesize{\scriptsize}
\rotate
\tablecaption{ Summary of the high-energy \textit{Fermi}-LAT power-law extrapolation beyond 150 GeV resulting from power-law fits from the 11 months of data used to select the candidates as well as the extrapolation based on the extended data set spanning 29 months.  These extrapolated flux values factor in the gamma-ray absorption resulting from interaction with the EBL and are reported in \% Crab Nebula flux units above the same energy threshold in order to allow direct comparison with the upper limits derived from the VERITAS observations.  \label{extrap}}
\tablewidth{0pt}
\tablehead{
 \colhead{Counterpart} & \colhead{Original $z$} & \colhead{1FGL}& \colhead{1FGL Integral Flux\tablenotemark{\dagger}}& \colhead{1FGL Extrapolated Flux} &
\colhead{Updated $z$} & \colhead{Updated Extrapolated Flux\tablenotemark{*}}  & \colhead{VERITAS UL} \\
 \colhead{Name} & \colhead{Used for Selection} & \colhead{Index\tablenotemark{\dagger}}& \colhead{$\ge$300 MeV}& \colhead{$\ge$ 150 GeV } &
\colhead{Lower Limit} & \colhead{[\%Crab]}  & \colhead{[\%Crab]} \\
\colhead{}  & \colhead{} &\colhead{}& \colhead{[$\times10^{-9}$ ph cm$^{-2}$s$^{-1}$]}&\colhead{[\% Crab] } &
\colhead{} &\colhead{} & \colhead{} }
\startdata
 RGB\,J0136+391 & 0.2 &1.73& 9.5&4.1&0.2&3.5 & 1.7 \\
 RGB\,J0316+090 & 0.2&1.72& 19.6&8.1&0.2& 0.3 &  2.0 \\
 RGB\,J0909+231  & 0.2&1.46&6.6&12.1&$\ge 0.4305$&0.5&  1.5 \\
 RGB\,J1058+564 & 0.143&1.97&15.7&1.5&0.143&1.1 &1.1 \\
 RGB\,J1243+364 &  0.2 &1.74&5.8&2.2&$\ge0.485$&0.8&  3.1 \\
 RX\,J1436.9+5639 &  0.15&1.45& 3.6&11.7&0.15&0.7& 2.4 \\
\enddata
\tablenotetext{\dagger}{Taken from \cite{abdo1FGL}. Index and flux extrapolated without error for target selection.}
\tablenotetext{*}{Computed with power-law fit from 29 months of data for steady sources and VERITAS coincident time window for variable sources (RGB\,J0316+090 and RGB\,J1058+564).}
\end{deluxetable}

\clearpage

\begin{deluxetable}{cccccccccccc}
\tabletypesize{\tiny}
\rotate
\tablecaption{ Summary of VERITAS observations and analysis results.  The significances are computed from counts extracted from source and background regions (ON and OFF, respectively) according to Equation 17 of \cite{lima}. The VHE 99\% confidence level integral upper limits (ULs) are used for the modeling and are derived with an assumed photon index of 3.0.  The percentage Crab Nebula flux values are given above the corresponding energy thresholds for each observation.\label{tbl-1}}
\tablewidth{0pt}
\tablehead{
 \colhead{Counterpart}  & \colhead{$z$} & \colhead{VERITAS} & \colhead{Observation}  & \colhead{ON} & \colhead{OFF} &\colhead{$\alpha$}\tablenotemark{\ddagger} & \colhead{Significance} & \colhead{Threshold\tablenotemark{\dagger}}& \colhead{Integral UL} & \colhead{Decorrelation\tablenotemark{\diamondsuit}}& \colhead{Differential UL} \\
\colhead{Name} & \colhead{}  & \colhead{Livetime [hr]} & \colhead{Window [MJD]}  &\colhead{Events} &\colhead{Events} &\colhead{} &\colhead{$\sigma$} & \colhead{Energy}& \colhead{Above Threshold} & \colhead{Energy [GeV]} & \colhead{at Decorrelation Energy} \\
\colhead{} & \colhead{}  & \colhead{} & \colhead{}  &\colhead{} &\colhead{} &\colhead{} &\colhead{} & \colhead{[GeV]}& \colhead{[$\times10^{-8}$ph m$^{-2}$s$^{-1}$] (\% Crab)} & \colhead{} & \colhead{[$\times10^{-7}$ m$^{-2}$s$^{-1}$TeV$^{-1}$]} }
\startdata
 RGB\,J0136+391  & -- & 9.9 & 55122-55129    & 1422 &11224 & 0.1277 & -0.2 &  165 &    $4.9$ (1.7\%)& 260&$1.6$\\
 RGB\,J0316+090  & -- & 4.4 & 55126-55133    & 698 &5560 &0.1282&-0.5&  165 &  $5.7$ (2.0\%) & 275 & 1.5\\
 RGB\,J0909+231  & $\ge$0.4305 & 14.2 & 55150-55204& 2141 &14310&0.1499 & -0.1 &165&  $4.3$ (1.5\%) &260&1.3\\
 RGB\,J1058+564  & 0.143 & 9.8 & 55160-55185    &    1415&10405&0.1417&-1.1 &   220 &$2.1$ (1.1\%)&350&4.7\\
 RGB\,J1243+364 & $\ge$0.485 & 11.5 & 55189-55291     &  1617&11133&0.1389& 0.9&   150 &9.9 (3.1\%)&250&2.9\\
 RX\,J1436.9+5639  & 0.15\tablenotemark{*} & 13.0 & 55275-55383      &  1563&12314&0.125& 0.6 & 240 & $4.1$ (2.4\%)&375&8.9\\
\enddata
\tablenotetext{\dagger}{Defined as the energy at which the differential rate of reconstructed gamma rays from the postulated source reaches its maximum, accurate to within the 20\% energy resolution of VERITAS.} 
\tablenotetext{\ddagger}{Normalization factor between size of the ON and OFF regions.}
\tablenotetext{\diamondsuit}{The energy at which the upper limit is minimally dependent on the index assumed for the source.}
\tablenotetext{*}{Redshift of spatially coincident galactic cluster, taken from \cite{bauer}.}
\end{deluxetable}

\clearpage

\begin{deluxetable}{cccccccc}
\tabletypesize{\scriptsize}
\rotate
\tablecaption{ Summary of \textit{Fermi}-LAT observations and analysis results.  A variability study was completed using 29 months of data for all sources.  For steady sources, the spectral analysis is completed for 29 months of data. The two variable sources RGB\,J0316+090 and RGB\,J1058+564 show spectral analysis results for LAT data coincident with a slightly expanded window as compared to the VERITAS observation window so as to allow a 5$\sigma$ detection.   \label{tbl-1}}
\tablewidth{0pt}
\tablehead{
 \colhead{Counterpart} &\colhead{1FGL} & \colhead{Variability Detected?} & \colhead{Prob\tablenotemark{*}} &
\colhead{MJD Fit} & \colhead{TS} & \colhead{Index} & \colhead{Integral Flux $\ge$ 300 MeV} \\
\colhead{Name} &\colhead{Name} & \colhead{} & \colhead{} &
\colhead{Window} &\colhead{} & \colhead{$\Gamma$} & \colhead{[$\times 10^{-9}$ ph cm$^{-2}$s$^{-1} $]}}
\startdata
 RGB\,J0136+391 &  J0136.5+3905 & no & 0.39 &  54682.7-55565.0   & 1430 & 1.78$\pm$0.04  &  12.0$\pm$0.7   \\
 RGB\,J0316+090 & J0316.1+0904 & yes & 0.00 & 55055.0-55145.0     & 45.7 &  2.09$\pm$0.26 &  9.5$\pm$3.3    \\
 RGB\,J0909+231 & J0909.2+2310 & no & 0.06 &   54682.7-55565.0  &123 &  1.68$\pm$0.13 &  1.8$\pm$0.5   \\
 RGB\,J1058+564 & J1058.6+5628 & yes  & 0.00 &   55160.0-55185.0  &37.2 &  1.98$\pm$0.24 &  12.4$\pm$3.8  \\
 RGB\,J1243+364 & J1243.1+3627 & no   & 0.60 &   54682.7-55565.0  & 627 & 1.76$\pm$0.06 &  5.6$\pm$0.5    \\
 RX\,J1436.9+5639 & J1437.0+5640 & no & 0.24 &   54682.7-55565.0  & 132 & 1.74$\pm$0.13 &  1.6$\pm$0.4     \\
\enddata
\tablenotetext{*}{Prob is the probability of steady emission as measured from $\Delta$TS per degree of freedom.}
\end{deluxetable}

\clearpage

\begin{deluxetable}{ccccccccc}
\tabletypesize{\scriptsize}
\rotate
\tablecaption{ Summary of \textit{Swift}-XRT observations and analysis results.  Photon counting mode data were fit with absorbed power laws using HI column densities from \cite{kalberla}.  Unabsorbed data were used for SED modeling.  For multiple observations showing no significant variability, exposures were combined to improve statistics.  If variability was detected between observations, both results are shown on the SED, although the model is only shown for the XRT observation falling closest to the window of VERITAS observations.\label{tbl-1}}
\tablewidth{0pt}
\tablehead{
 \colhead{Counterpart} &\colhead{Observation} & \colhead{MJD} & \colhead{Exposure} &\colhead{HI} & \colhead{Flux (2-10 keV)} & \colhead{Photon} & \colhead{$\chi^2$ (DOF)} & \colhead{Used in} \\
\colhead{Name} & \colhead{ID} &\colhead{} &\colhead{Time [ks]} & \colhead{[$\times10^{20}$cm$^{-2}$]} & \colhead{[$\times10^{-12}$ ergs cm$^{-2}$s$^{-1}$]} &
\colhead{Index} & \colhead{} & \colhead{SED} \\}
\startdata
 RGB\,J0136+391 & 00039107001 & 55134 & 1.6  &6.17 &12.1$\pm$1.0  & 2.24$\pm$0.05      &  38.0 (40) & yes  \\
                              & 00039107002 & 55223 &  2.9 &6.17 & 21.4$\pm$1.0 & 2.16$\pm$0.03      &107.2 (84) & yes \\
 \hline
 RGB\,J0316+090 & 00038370001 & 54899 &2.6 & 12.6& 7.0$\pm$0.6  &2.13$\pm$0.06   &34.7 (35) & yes \\
 \hline
 RGB\,J0909+231\tablenotemark{*} & 00040540001 & 55338 & 1.3&$--$& $--$  &$--$   & $--$ & no\\
                              & 00040540002 & 55338 & 1.9&$--$&  $--$  &  $--$ &$--$&  no\\
                              & 00040540003 & 55342& 1.6 &$--$&  $--$  &  $--$ &$--$ & no\\   
                              & combined       &$--$ & $--$&4.63&0.3$\pm$0.06&2.7$\pm$0.2&5.1 (5)  & yes \\
 \hline
 RGB\,J1058+564 & 00038215001 & 54852 & 3.8&0.78 &2.2$\pm$0.2 &2.60$\pm$0.05    &  42.4 (39) & yes\\
                              & 00038453001 & 54888  &1.0&0.78 &6.3$\pm$0.6 &2.48$\pm$0.07   &   25.9 (27)& yes\\
 \hline
 RGB\,J1243+364 & 00038445001 & 54875 &2.1&1.4 & 8.2$\pm$0.5 &2.33$\pm$0.05   & 50.4 (46) & yes\\
 \hline
 RX\,J1436.9+5639 & 00038435001 & 54918 &5.1 &1.55& 2.4$\pm$0.2   &2.31$\pm$0.05  &40.6 (31)  & no \\
                                & 00038289001 & 54918  &4.9&1.55&   2.4$\pm$0.2  &2.27$\pm$0.05    &43.3 (31) & no\\
                                & combined        & $--$&$--$ &1.55 &2.4$\pm$0.2 &2.29$\pm$0.04  &72.5 (78) & yes\\
\enddata
\tablenotetext{*}{Statistics extracted from each single observation are too low to fit an absorbed power law.  The combination of the three exposures is used.}
\end{deluxetable}

\clearpage

\begin{deluxetable}{ccccccc}
\tabletypesize{\scriptsize}
\tablecaption{Summary of \textit{Swift}-UVOT observations and analysis results. \label{tbl-1}}
\tablewidth{0pt}
\tablehead{
 \colhead{Target} & \colhead{Observation} &\colhead{Date} &\colhead{Band} & \colhead{Frequency} & \colhead{$\nu$F$_{\nu}$}  & \colhead{Used in}\\
 \colhead{} & \colhead{ID} & \colhead{[MJD]}& \colhead{} & \colhead{[Hz]} & \colhead{[Jy Hz]} & \colhead{SED?} \\}
\startdata
 RGB\,J0136+391 & 00039107001 & 55134 &UVW1  &$1.14\times10^{15}$ &$(1.16\pm0.15)\times10^{12}$ & y \\
                              & 00039107001 & 55134& UVM2 & $1.34\times10^{15}$&$(1.44\pm0.30)\times10^{12}$  & y \\
 \hline
 RGB\,J0316+090 & 00038370001 & 54899 & V & $5.55\times10^{14}$ &$(2.24\pm1.04)\times10^{12}$&y\\
                              & 00038370001 & 54899 & U & $8.57\times10^{14}$ &$(2.92\pm3.87)\times10^{12}$&n\tablenotemark{*}\\
                              & 00038370001 & 54899 & UVW1 & $1.14\times10^{15}$&$(2.44\pm7.64)\times10^{12}$&n\tablenotemark{*}\\
                              & 00038370001 & 54899 & UVW2 &$1.48\times10^{15}$ &$(2.8\pm18.5)\times10^{12}$&n\tablenotemark{*}\\
 \hline
 RGB\,J0909+231 & 00040540001 & 55338  & V & $5.55\times10^{14}$&$(2.35\pm0.11)\times10^{11}$&y\\
                              & 00040540001 & 55338  & B & $6.93\times10^{14}$&$(1.77\pm0.23)\times10^{11}$&y\\
                              & 00040540001 & 55338  & U &$8.57\times10^{14}$ &$(2.26\pm0.21)\times10^{11}$&y\\   
                              & 00040540001 & 55338  & UVW1 &$1.14\times10^{15}$&$(1.78\pm0.18)\times10^{11}$ &y\\
                              & 00040540001 & 55338  & UVM2 &$1.34\times10^{15}$ &$(1.63\pm0.22)\times10^{11}$&y\\
                              & 00040540001 & 55338  & UVW2 &$1.48\times10^{15}$ &$(1.80\pm0.19)\times10^{11}$&y\\  
                              & 00040540002 & 55338  & UVM2 & $1.34\times10^{15}$&$(1.90\pm0.22)\times10^{11}$&n\\    
                              & 00040540003 & 55338  & UVW2 & $1.48\times10^{15}$&$(1.74\pm0.20)\times10^{11}$&n\\    
 \hline
 RGB\,J1058+564 & 00038215001 & 54852 & V &$5.55\times10^{14}$ &$(1.42\pm0.04)\times10^{12}$&y\\
                               & 00038215001 & 54852 & U &$8.57\times10^{14}$ &$(1.45\pm0.04)\times10^{12}$&y\\
                               & 00038215001 & 54852 & UVW2 &$1.48\times10^{15}$ &$(1.49\pm0.05)\times10^{12}$&y\\
                              & 00038453001 & 54888 & B &$6.93\times10^{14}$ &$(1.64\pm0.07)\times10^{12}$&n\\
                              & 00038453001 & 54888 & U & $8.57\times10^{14}$&$(1.70\pm0.06)\times10^{12}$&n\\                              
                              & 00038453001 & 54888 & UVW1 & $1.14\times10^{15}$&$(1.66\pm0.06)\times10^{12}$&n\\
 \hline
 RGB\,J1243+364 & 00038445001 & 54875& V &$5.55\times10^{14}$ &$(1.28\pm0.05)\times10^{12}$ &y \\
                              & 00038445001 & 54875& B & $6.93\times10^{14}$&$(1.30\pm0.04)\times10^{12}$ & y\\
                              & 00038445001 & 54875& U & $8.57\times10^{14}$&$(1.42\pm0.05)\times10^{12}$ & y\\
                              & 00038445001 & 54875& UVW1 & $1.14\times10^{15}$&$(1.34\pm0.05)\times10^{12}$ & y\\
                              & 00038445001 & 54875& UVM2 &$1.34\times10^{15}$ & $(1.49\pm0.05)\times10^{12}$& y\\
                              & 00038445001 & 54875& UVW2 & $1.48\times10^{15}$& $(1.47\pm0.05)\times10^{12}$&y \\
 \hline
 RX\,J1436.9+5639 & 00038435001 & 54918  & V &$5.55\times10^{14}$ &$(9.03\pm0.09)\times10^{11}$ & y\\
                                & 00038435001 & 54918  & B & $6.93\times10^{14}$&$(1.14\pm0.07)\times10^{12}$ & y\\
                                & 00038435001 & 54918  & U &$8.57\times10^{14}$ &$(9.71\pm0.05)\times10^{11}$ & y\\
                                & 00038435001 & 54918  & UVW1 &$1.14\times10^{15}$ & $(1.06\pm0.06)\times10^{12}$& y\\                                
                                & 00038435001 & 54918  & UVM2 & $1.34\times10^{15}$&$(1.21\pm0.07)\times10^{12}$ & y\\
                                & 00038435001 & 54918  & UVW2 &$1.48\times10^{15}$ & $(1.12\pm0.06)\times10^{12}$& y\\
                                & 00038289001 & 54918 & V & $5.55\times10^{14}$&$(8.58\pm0.05)\times10^{11}$&n\\
                                & 00038289001 & 54918 & U & $8.57\times10^{14}$&$(1.06\pm0.04)\times10^{12}$&n\\
                                & 00038289001 & 54918 & UVW2 &$1.48\times10^{15}$ &$(1.26\pm0.06)\times10^{12}$&n\\

\enddata
\tablenotetext{*}{Galactic reddening dominates the uncertainty with E(B-V) = 0.356 for RGB\,J0316+090; U, UVW1, UVW2 band flux measurements do not constrain the SED.}
\end{deluxetable}

\clearpage

\begin{deluxetable}{ccccccccccccc}
\tabletypesize{\scriptsize}
\rotate
\tablecaption{Summary of the broadband SED SSC modeling parameters.  See text for parameter descriptions.\label{tbl-5}}
\tablewidth{0pt}
\tablehead{
 \colhead{Counterpart Name} & \colhead{$z$}&\colhead{$\gamma_{min}$\tablenotemark{a}} & \colhead{$\gamma_{max}$\tablenotemark{a}} & \colhead{$q$\tablenotemark{a}} &
\colhead{$t_{esc}$ [hr]\tablenotemark{a}} & \colhead{B [Gauss]\tablenotemark{a}} & \colhead{$\Gamma$\tablenotemark{a}} & \colhead{Radius [cm]\tablenotemark{a}}  & \colhead{$L_e$ [ergs/sec]\tablenotemark{a}}& \colhead{$L_B$ [ergs/sec]\tablenotemark{b}}&\colhead{$L_B/L_e$\tablenotemark{b}} &\colhead{$t_{min}$[hr]\tablenotemark{b}}}
\startdata
 RGB\,J0136+391  &0.2\tablenotemark{*}      & $4.0\times10^4$ & $6.0\times10^5$ & 2.6 & 120 &  0.15 &15  &$3.4\times10^{16}$ &$1.1\times10^{44}$ &$2.3\times10^{43}$ & $2.1\times10^{-1}$&25.9\\
 
 RGB\,J0136+391  &0.3\tablenotemark{*}      & $2.5\times10^4$ & $5.0\times10^5$ & 2.6 &   50 &  0.4   &15  &$2.4\times10^{16}$ &$1.0\times10^{44}$ &$7.8\times10^{43}$ &$7.5\times10^{-1}$ &19.2\\
 
 RGB\,J0136+391  &0.4\tablenotemark{*}      & $2.4\times10^4$ & $5.0\times10^5$ & 2.6 &   30 &  0.45 &15  &$3.0\times10^{16}$ & $1.3\times10^{44}$& $1.5\times10^{44}$&1.1 &25.9\\
 
 RGB\,J0316+090  &0.2\tablenotemark{*}      & $1.5\times10^4$ & $7.0\times10^5$ & 2.5 &   15 &  0.11 &15  &$8.0\times10^{16}$ & $9.8\times10^{43}$& $6.5\times10^{43}$ &$6.6\times10^{-1}$ &59.2\\
 
 RGB\,J0909+231  &0.5      & $2.3\times10^4$ & $5.0\times10^5$ & 2.8 &   10 &  0.1   & 20 &$2.5\times10^{16}$ &$1.5\times10^{44}$ & $9.4\times10^{42}$&$6.4\times10^{-2}$ &17.4\\
 
 RGB\,J1058+564  &0.1433& $4.0\times10^4$ & $3.0\times10^5$ & 2.9 &   10 &  0.25 & 20 &$9.0\times10^{15}$ &$3.9\times10^{43}$ & $7.6\times10^{42}$& $2.0\times10^{-1}$&4.8\\
 
 RGB\,J1243+364  &0.5      & $3.0\times10^4$ & $1.0\times10^6$ & 2.5 &   30 &  0.1   & 20 &$1.5\times10^{17}$ &$3.2\times10^{44}$ & $3.4\times10^{44}$& 1.1&104.1\\
 
 RX\,J1436.9+5639            & 0.15   & $4.0\times10^4$ & $2.5\times10^6$ & 2.6 & 100 &  0.01 & 20 &$2.7\times10^{17}$ &$1.5\times10^{44}$ &$1.1\times10^{43}$ &$7.5\times10^{-2}$&143.7\\
\enddata
\tablenotetext{*}{Assumed redshift values due to lack of any redshift constraining measurement.}
\tablenotetext{a}{Parameters adjusted to match the model to the data.}
\tablenotetext{b}{Quantities derived from the modeling result.}
\end{deluxetable}

\clearpage

\end{document}